\documentclass[journal]{IEEEtran}
\usepackage{amsmath}
\usepackage{graphicx}
\ifCLASSINFOpdf
\else
\fi
\usepackage{caption}
\usepackage{soul}
\usepackage{bm}
\usepackage{amssymb}
\usepackage{algorithm}
\usepackage{algpseudocode}
\usepackage{multirow}
\usepackage{xcolor}
\usepackage[inline]{enumitem}
\usepackage{mathtools}
\definecolor{steelblue}{RGB}{48 126 167}
\usepackage{varioref}
\usepackage{url}
\usepackage{hyperref}
\hypersetup{
  colorlinks = true,   
  linkcolor =  black,
  urlcolor =   steelblue,
  citecolor =  black,
  plainpages =        false,
  hypertexnames =     true,
  linktocpage =       true,
  bookmarksopen =     true,
  bookmarksnumbered = true,
  bookmarksopenlevel= 0,
}
\usepackage[all]{hypcap}
\usepackage[noabbrev,capitalise,nameinlink]{cleveref}
\definecolor{lightorange}{RGB}{255, 204, 153}
\usepackage{cite}
\begin{document}
%
\title{Segmentation-Guided Knee Radiograph Generation using Conditional Diffusion Models}
%
%
%

\author{Siyuan~Mei,
        Fuxin~Fan,
        Fabian~Wagner,
        Mareike~Thies,
        Mingxuan~Gu,
        Yipeng Sun,
        and~Andreas~Maier
\thanks{All authors are with the Pattern Recognition Lab, Friedrich-Alexander-Universität Erlangen-Nürnberg, Germany. Contact E-mail: siyuan.mei@fau.de.}}
\maketitle

\begin{abstract}
Deep learning-based medical image processing algorithms require representative data during development. In particular, surgical data might be difficult to obtain, and high-quality public datasets are limited. To overcome this limitation and augment datasets, a widely adopted solution is the generation of synthetic images. In this work, we employ conditional diffusion models to generate knee radiographs from contour and bone segmentations. Remarkably, two distinct strategies are presented by incorporating the segmentation as a condition into the sampling and training process, namely, conditional sampling and conditional training. The results demonstrate that both methods can generate realistic images while adhering to the conditioning segmentation. The conditional training method outperforms the conditional sampling method and the conventional U-Net.

\end{abstract}


%
\IEEEpeerreviewmaketitle

\section{Introduction}
%
%
%
%
\IEEEPARstart{R}{adiography} is one of the most commonly used medical imaging techniques for diagnosis and surgical interventions, capturing 2D projection images of patients using X-ray. The advent of deep learning-based techniques has enabled automated and precise processing of X-ray images, including organ segmentation~\cite{seg2}, motion compensation~\cite{thies2023gradient}, and denoising~\cite{wagner2022ultralow}. However, many of these methods heavily rely on extensive training data which is challenging to collect at scale~\cite{dataaugmentation}. In particular, recent research demands atypical data, such as weight-bearing imaging of knees~\cite{bier2018detecting, choi2013fiducial}. 

To address this challenge, recent works proposed synthesizing simulated data as a substitute for clinical data~\cite{gao2023synthetic}. Traditional forward projection methods create digitally reconstructed radiographs (DRRs) from 3D computed tomography (CT) volumes using Radon transform~\cite{kak2001principles}, which guarantees geometrical accuracy. However, such methods require original volumetric CT scans, and sophisticated forward models are needed to capture all properties of realistic-looking X-rays (e.g., energy-dependent effects, noise, scatter, etc.). In contrast, deep generative models leverage 2D X-ray datasets to generate radiographs. For instance, Weber et al.~\cite{weber2022implicit} employed generative adversarial network (GAN)-based models to augment chest X-ray images. 


Recently, diffusion models have emerged as a powerful technique for data generation, demonstrating competitive performance compared to GANs~\cite{diffusionbeatgan}. In addition, diffusion models are also successfully applied in conditional radiograph generation, e.g., projection inpainting conditioned on masked projection~\cite{mei2023metal} and class-conditional chest radiograph synthesis~\cite{latent}. Despite achieving convincing performance, previous work has mainly focused on generating radiographs under specific conditions. In this work, we focus on knee imaging and extend the conditional generation of knee radiographs to include more general conditions, such as simple segmentation with contour and bone information. To learn the conditional distribution given the segmentation, we propose two distinct pipelines of conditional diffusion models that incorporate conditional images into the sampling and training processes, respectively.


\section{Methods and Materials}

\subsection{Method of Conditional Sampling}

\begin{figure*}[htb]
    \begin{center}
    \begin{tabular}{c} 
    \includegraphics[width=12cm]{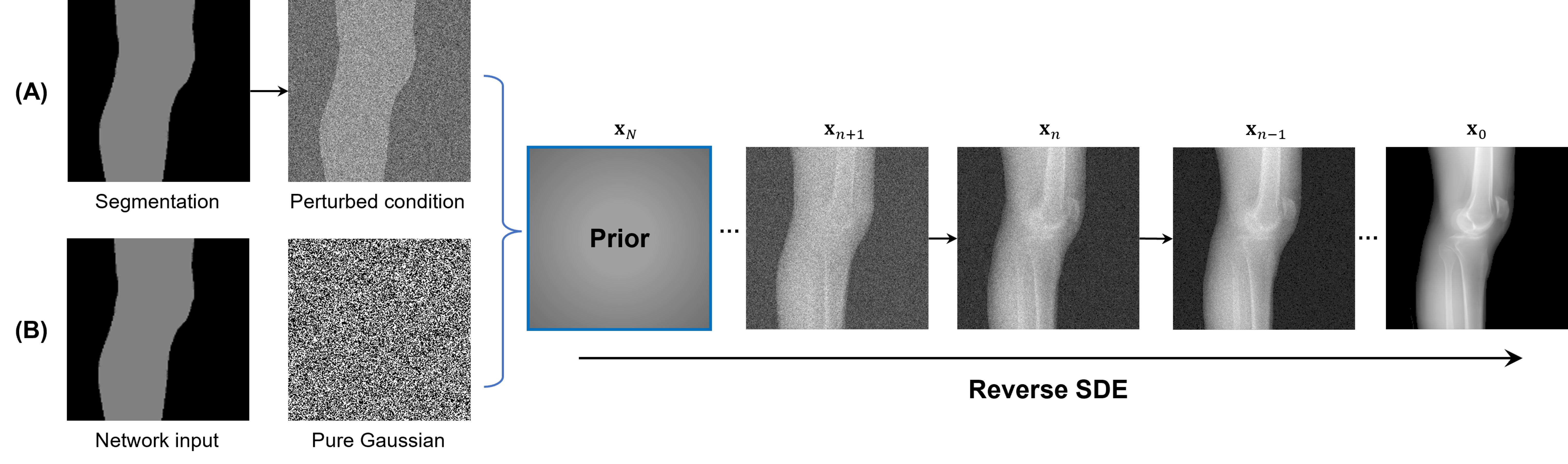}
    \end{tabular}
    \end{center}
    \captionsetup{justification=centering}
    
    \caption
    {Overview of the sampling processes. (A) illustrates the sampling process of the conditional sampling method; (B) illustrates the sampling process of the conditional training method.}
    \label{fig:figure1}
\end{figure*}

Diffusion models employ a forward diffusion process and a reverse diffusion process for image generation. In the forward process, data points gradually diffuse into random Gaussian noise over time, implying the transition from a complex to a simple data distribution. Conversely, the reverse diffusion process generates new data samples by progressively removing noise, starting from a Gaussian prior. Elegantly, the forward perturbation can be modeled as a stochastic differential equation (SDE), which is tractable throughout the reverse process~\cite{song2020score}. 

Let $\mathbf{x}_0\in\mathbb{R}^d$ denote the data sample, $\mathbf{x}_t$ denote the perturbed data at time point $t\in (0, 1]$, and $p(\mathbf{x}_t)$ denote the corresponding probability density function of $\mathbf{x}_t$. The forward SDE is defined by
\begin{equation}
    {\rm d}\mathbf{x}_t = \mathbf{f}(\mathbf{x}_t,t){\rm d}t+g(t)\mathbf{z}_t,\quad t:0\rightarrow 1,
    \label{forward sde}
\end{equation}
where $\mathbf{f}(\mathbf{x}_t,t)\in\mathbb{R}^d$ and $g(t)\in \mathbb{R}$ are the drift and diffusion coefficients of $\mathbf{x}_t$, and $\mathbf{z}_t\sim\mathcal{N}(0, \mathbf{I})$ defines a random variable drawn from a standard normal distribution approximating the standard Wiener process. The corresponding reverse-time SDE has the form 
\begin{equation}
    {\rm d}\overline{\mathbf{x}}_t = [\mathbf{f}(\mathbf{x}_t,t)-g(t)^2\nabla _{\mathbf{x}_t} \log p(\mathbf{x}_t)]{\rm d}t+g(t)\mathbf{z}_t,\quad t:1\rightarrow 0,
    \label{reverse sde}
\end{equation}
where $\nabla _{\mathbf{x}_t} \log p(\mathbf{x}_t)$ represents the gradient of the logarithmic probability density, also known as the score function.

In this paper, we adopt the variance exploding (VE) SDE setting~\cite{song2020score}, defining the growing standard deviation of the noise as
\begin{equation}
\sigma_t=\sigma_{min}\left(\frac{\sigma_{max}}{\sigma_{min}}\right)^t, 
\end{equation}
which determines the drift and diffusion coefficients in~\eqref{forward sde} and~\eqref{reverse sde} as
\begin{equation}
    \mathbf{f}(\mathbf{x}_t,t)=0,\quad g(t)=\sigma_t\sqrt{2\log \frac{\sigma_{max}}{\sigma_{min}}}.
    \label{eq: ve sde}
\end{equation}

Crucially, the last unknown term $\nabla _{\mathbf{x}_t} \log p(\mathbf{x}_t)$ in~\eqref{reverse sde} can be estimated by training a time-conditional neural network $\mathbf{s}_{\bm \theta}(\mathbf{x}_t,t)$. The optimal parameter ${\bm \theta}^\ast$ is obtained by minimizing denoising score matching~\cite{score_matching}
\begin{align}\nonumber
   {\bm \theta}^\ast&=\mathop{{\rm argmin}}_{{\bm \theta}}\mathbb{E}_{t\sim U[0,1]}\left[\lVert \mathbf{s}_{\bm \theta} (\mathbf{x}_t, t)-\nabla _{\mathbf{x}_t} \log p(\mathbf{x}_t)\rVert_2^2\right]\\
    &\approx \mathop{{\rm argmin}}_{{\bm \theta}}\mathbb{E}_{t\sim U[0,1]}\left[\lVert \mathbf{s}_{\bm \theta} (\mathbf{x}_t, t)-\nabla _{\mathbf{x}_t} \log p(\mathbf{x}_t|\mathbf{x}_0)\rVert_2^2\right],
\end{align}
where $p(\mathbf{x}_t|\mathbf{x}_0)$ is the perturbation kernel of the VE SDE.


With segmentation as known information (denoted as $\mathbf{y}$), the conditional sampling method (CSM) follows the SDEdit algorithm~\cite{meng2021sdedit} to synthesize knee radiographs. As depicted in Fig~\ref{fig:figure1} (A), CSM commences with a perturbed leg contour segmentation, and realistic details are generated through iterative denoising while retaining the desired shape. Notably, the initial perturbing noise is appropriately reduced to preserve conditional information by setting the starting time point $t_0$ of the reverse diffusion process smaller than 1. Therefore, the prior sample becomes 
\begin{equation}
    \mathbf{y}_{t_0} \sim \mathcal{N}(\mathbf{y},\sigma_{t_0}^2\mathbf{I}).
\end{equation}
The pipeline of CSM is described in algorithm~\ref{alg: conditional sampling}.
\begin{algorithm}
\caption{Sampling algorithm of CSM}
\label{alg: conditional sampling}
\begin{algorithmic}[1]
\Require $N$ (number of sampling steps), $\mathbf{y}$ (segmentation guide), $t_0$ (starting time of the reverse diffusion process) 
\State $\mathbf{x}_{N} \sim$ \colorbox{lightorange}{$\mathbf{y} + \mathcal{N}(0,\sigma_{t_0}^2\mathbf{I})$}
\For{$n = N-1$ to $0$}
    \State $\mathbf{z} \sim \mathcal{N}(0,\mathbf{I})$  
    \State $t_n \gets$ \colorbox{lightorange}{$\frac{n}{N} t_0$}  
    \State $\Delta\overline{\mathbf{x}}_n \gets g(t_n)^2$\colorbox{lightorange}{$\mathbf{s}_{{\bm \theta}^\ast}(\mathbf{x}_n,t_n)$}$+g(t_n)\mathbf{z}$
    \State $\mathbf{x}_n \gets \mathbf{x}_{n+1} + \Delta\overline{\mathbf{x}}_n$
    \EndFor
\State \Return $\mathbf{x}_0$
\end{algorithmic}
\end{algorithm}

\subsection{Method of Conditional Training}

An alternative to the CSM is integrating the conditions into the training process, thereby directly estimating the score function of the conditional distribution~\cite{saharia2022image}. We concatenate the condition and the perturbed image along the channel dimension as network input. The structure of the score-based network is detailed in section~\ref{sec: network}. Surprisingly, this conditional score network can be trained following the same form of denoising score matching~\cite{batzolis2021conditional}, which is
\begin{align}\nonumber
   {\bm \theta}^\ast&=\mathop{{\rm argmin}}_{{\bm \theta}}\mathbb{E}_{t\sim U[0,1]}\left[\lVert \mathbf{s}_{\bm \theta} (\mathbf{x}_t, \mathbf{y}, t)-\nabla _{\mathbf{x}_t} \log p(\mathbf{x}_t|\mathbf{y})\rVert_2^2\right]\\
    &\approx \mathop{{\rm argmin}}_{{\bm \theta}}\mathbb{E}_{t\sim U[0,1]}\left[\lVert \mathbf{s}_{\bm \theta} (\mathbf{x}_t, \mathbf{y}, t)-\nabla _{\mathbf{x}_t} \log p(\mathbf{x}_t|\mathbf{x}_0)\rVert_2^2\right].
\end{align}

After training the conditioned network, the score function in~\eqref{reverse sde} can be directly substituted by $\mathbf{s}_{\bm \theta^*} (\mathbf{x}_t, \mathbf{y}, t)$ for the provided conditional image in each sampling step. As shown in Fig.~\ref{fig:figure1} (B), the generated sample is controlled even though the sampling process starts from random Gaussian noise. Algorithm~\ref{alg: conditional training} outlines the sampling process of the conditional training method (CTM), where the modifications with respect to algorithm~\ref{alg: conditional sampling} are highlighted in orange.

\begin{algorithm}
\caption{Sampling algorithm of CTM}
\label{alg: conditional training}
\begin{algorithmic}[1]
\Require $N$ (number of sampling steps), $\mathbf{y}$ (segmentation guide)
\State $\mathbf{x}_N \sim$\colorbox{lightorange}{$\mathcal{N}(0,\sigma_{max}^2\mathbf{I})$}
\For{$n = N-1$ to $0$}
    \State $\mathbf{z} \sim \mathcal{N}(0,\mathbf{I})$  
    \State $t_n \gets $\colorbox{lightorange}{$\frac{n}{N}$}  
    \State $\Delta\overline{\mathbf{x}}_n \gets g(t_n)^2$\colorbox{lightorange}{$\mathbf{s}_{{\bm \theta}^\ast}(\mathbf{x}_n, \mathbf{y}, t_n)$}$+g(t_n)\mathbf{z}$
    \State $\mathbf{x}_n \gets \mathbf{x}_{n+1} + \Delta\overline{\mathbf{x}}_n$
    \EndFor
\State \Return $\mathbf{x}_0$
\end{algorithmic}
\end{algorithm}

\subsection{Dataset Preparation}

\begin{figure*}[htb]
    \begin{center}
    \begin{tabular}{c} 
    \includegraphics[width=14cm]{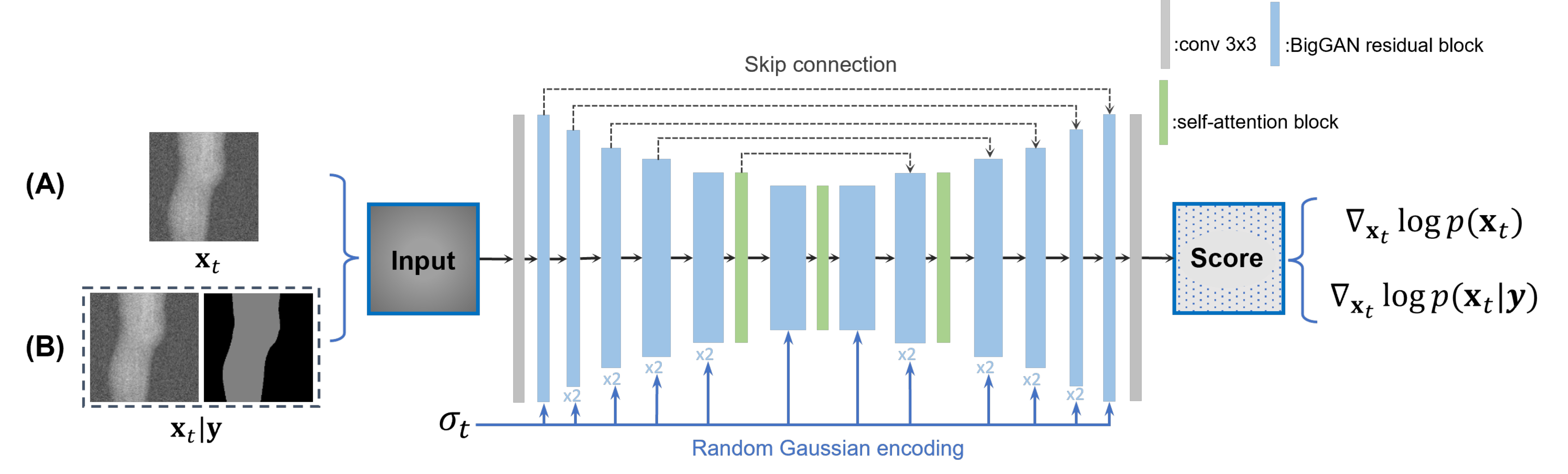}
    \end{tabular}
    \end{center}
    \captionsetup{justification=centering}
    \caption[figure2] 
    { \label{fig:figure2} 
    Structure of noise-conditional score network. (A) represents the network input of CSM; (B) illustrates the network input of CTM.}
\end{figure*}

To obtain knee radiographs, we selected 55 leg CT volumes from the public SICAS medical image repository~\cite{kistler2013virtual}. Each CT volume was simulated over $360^\circ$ with an angular increment of $6^\circ$ and projected onto the detector of the size $256\times 256$ using CONRAD~\cite{maier2013conrad}, resulting in 60 DRRs per volume. All projections were normalized to the range of [0,1]. 

\begin{figure}[htb!]
\centering

    \begin{minipage}[t]{0.18\linewidth}
    \centering
    \centerline{Condition}\medskip
\end{minipage}
\hfill
    \begin{minipage}[t]{0.18\linewidth}
    \centering
    \centerline{U-Net}\medskip
\end{minipage}
\hfill
    \begin{minipage}[t]{0.18\linewidth}
    \centering
    \centerline{CSM}\medskip
\end{minipage}
\hfill
    \begin{minipage}[t]{0.18\linewidth}
    \centering
    \centerline{CTM}\medskip
\end{minipage}
\hfill
    \begin{minipage}[t]{0.18\linewidth}
    \centering
    \centerline{Label}\medskip
\end{minipage}

    \begin{minipage}[b]{0.18\linewidth}
    \centering
    \centerline{\includegraphics[width=1.6cm]{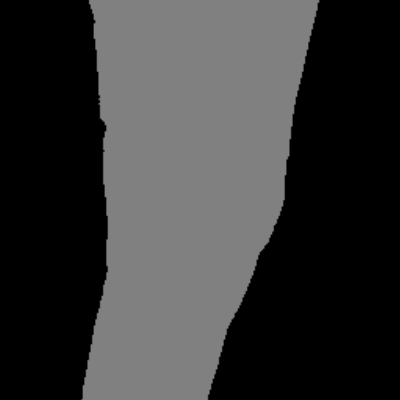}}
    \centerline{(a0)}\medskip
\end{minipage}
\hfill
    \begin{minipage}[b]{0.18\linewidth}
    \centering
    \centerline{\includegraphics[width=1.6cm]{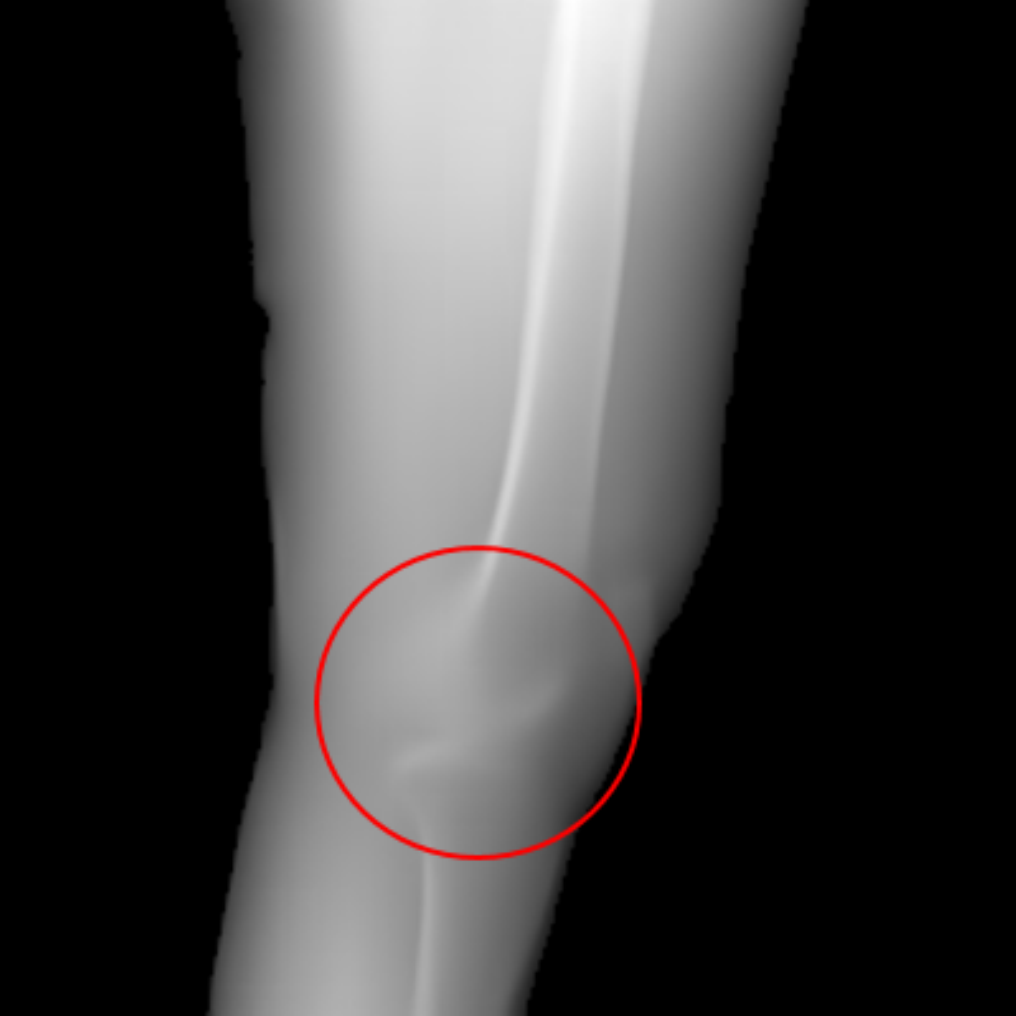}} 
    \centerline{(a1)}\medskip
\end{minipage}
\hfill
    \begin{minipage}[b]{0.18\linewidth}
    \centering
    \centerline{\includegraphics[width=1.6cm]{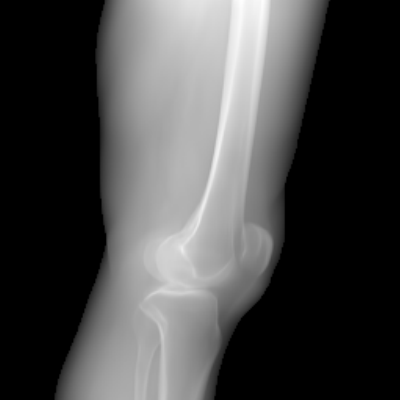}} 
    \centerline{(a2)}\medskip
\end{minipage}
\hfill
    \begin{minipage}[b]{0.18\linewidth}
    \centering
    \centerline{\includegraphics[width=1.6cm]{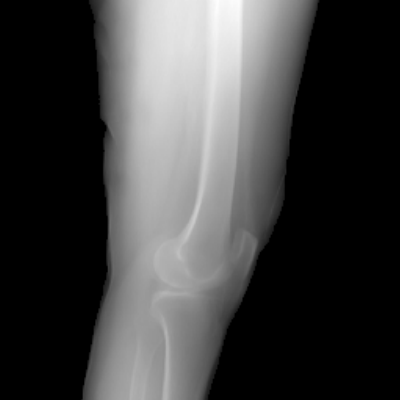}}
    \centerline{(a3)}\medskip
\end{minipage}
\hfill
    \begin{minipage}[b]{0.18\linewidth}
    \centering
    \centerline{\includegraphics[width=1.6cm]{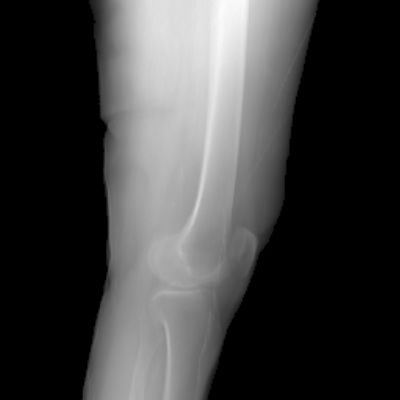}}
    \centerline{(a4)}\medskip
\end{minipage}

    \begin{minipage}[b]{0.18\linewidth}
    \centering
    \centerline{\includegraphics[width=1.6cm]{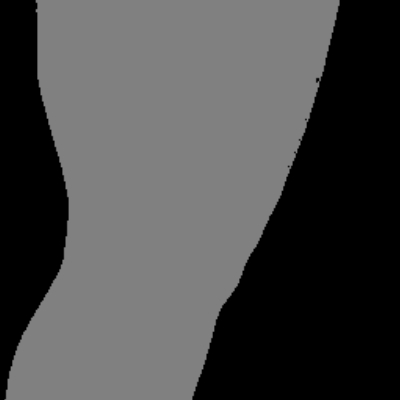}}
    \centerline{(b0)}\medskip
\end{minipage}
\hfill
    \begin{minipage}[b]{0.18\linewidth}
    \centering
    \centerline{\includegraphics[width=1.6cm]{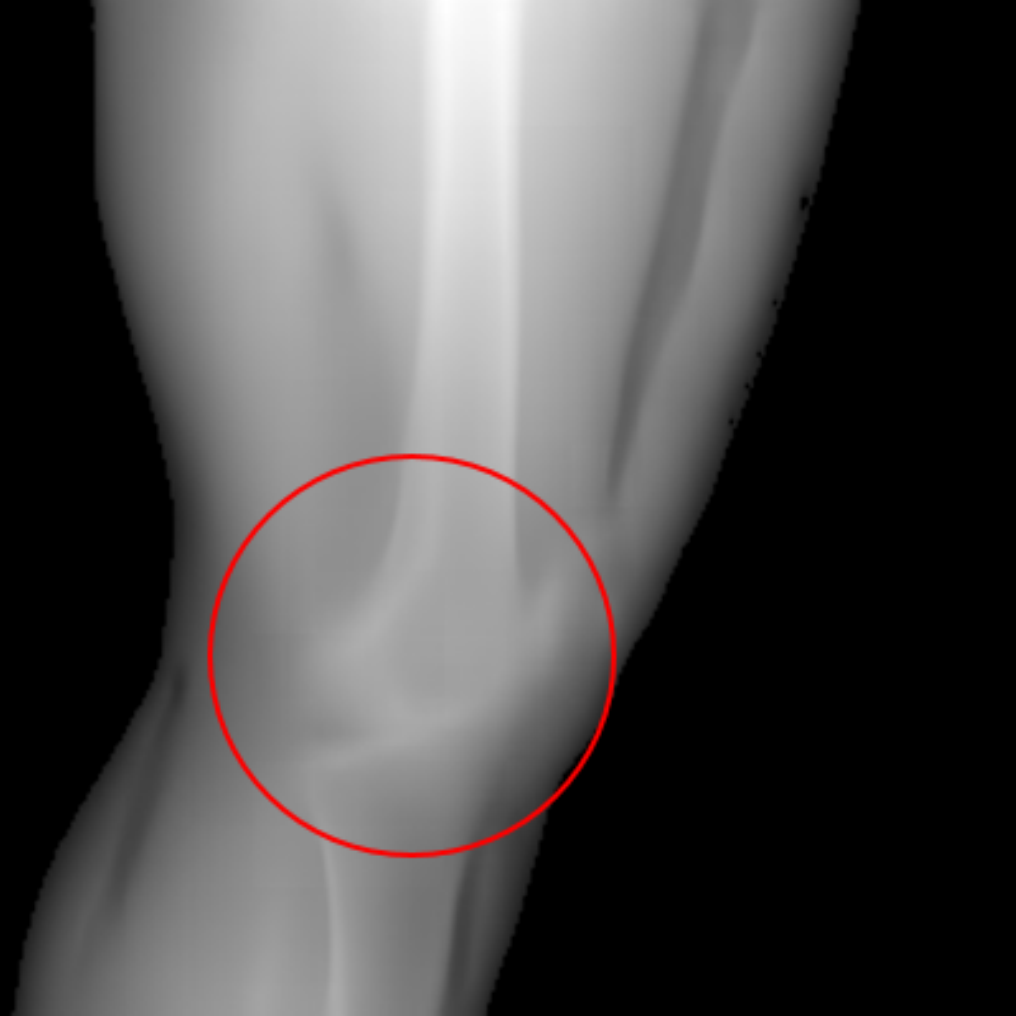}} 
    \centerline{(b1)}\medskip
\end{minipage}
\hfill
    \begin{minipage}[b]{0.18\linewidth}
    \centering
    \centerline{\includegraphics[width=1.6cm]{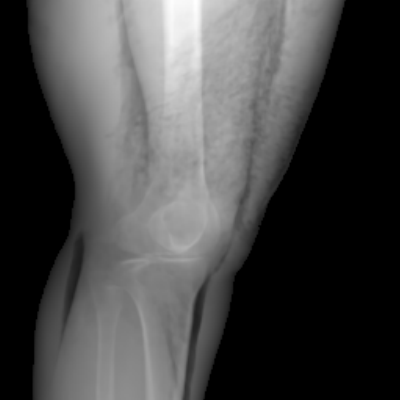}} 
    \centerline{(b2)}\medskip
\end{minipage}
\hfill
    \begin{minipage}[b]{0.18\linewidth}
    \centering
    \centerline{\includegraphics[width=1.6cm]{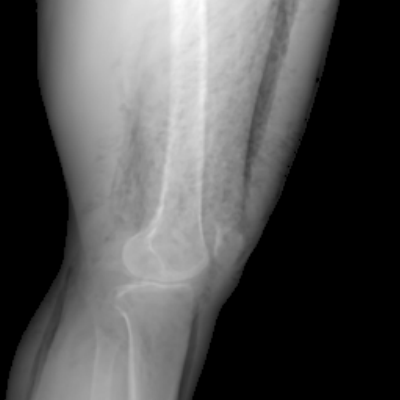}}
    \centerline{(b3)}\medskip
\end{minipage}
\hfill
    \begin{minipage}[b]{0.18\linewidth}
    \centering
    \centerline{\includegraphics[width=1.6cm]{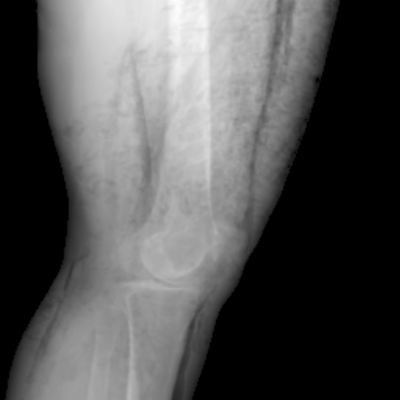}}
    \centerline{(b4)}\medskip
\end{minipage}

    \begin{minipage}[b]{0.18\linewidth}
    \centering
    \centerline{\includegraphics[width=1.6cm]{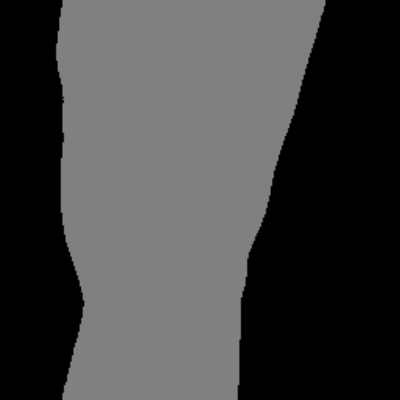}}
    \centerline{(c0)}\medskip
\end{minipage}
\hfill
    \begin{minipage}[b]{0.18\linewidth}
    \centering
    \centerline{\includegraphics[width=1.6cm]{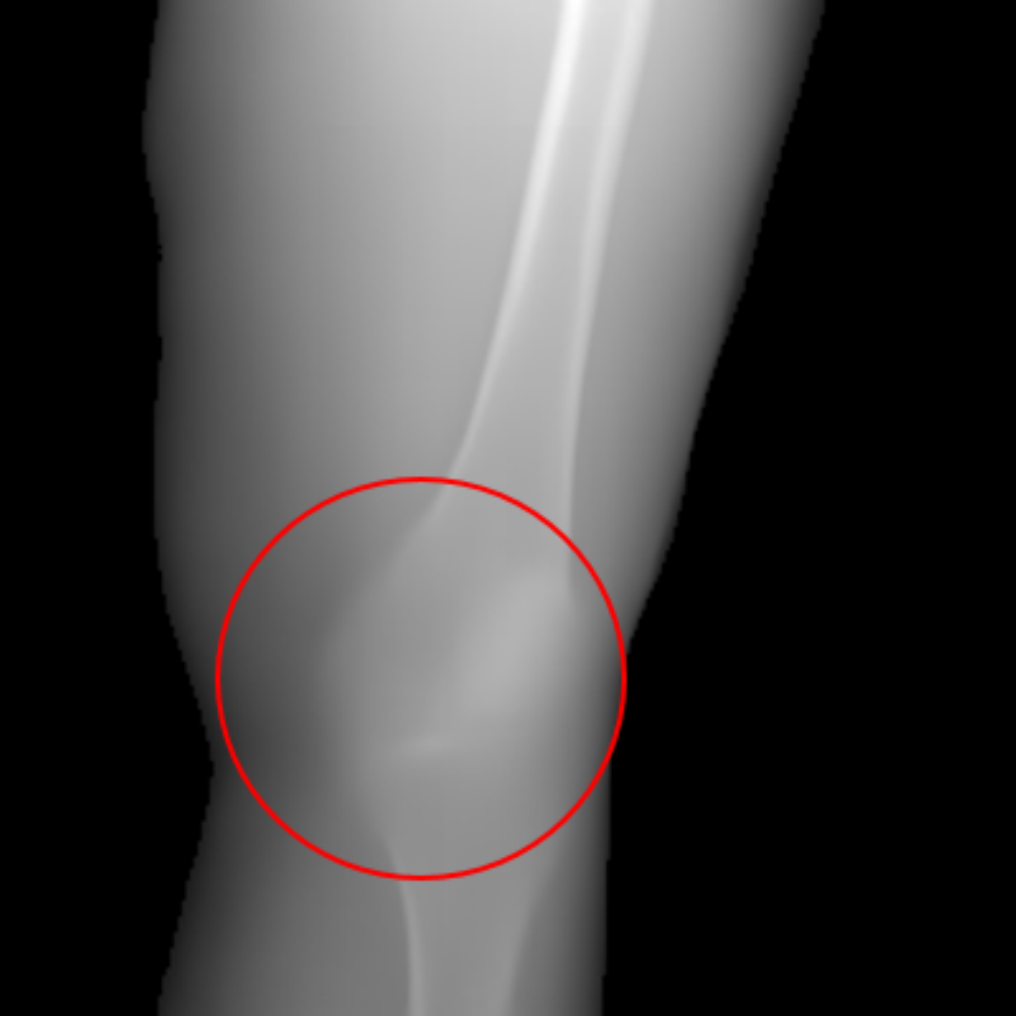}} 
    \centerline{(c1)}\medskip
\end{minipage}
\hfill
    \begin{minipage}[b]{0.18\linewidth}
    \centering
    \centerline{\includegraphics[width=1.6cm]{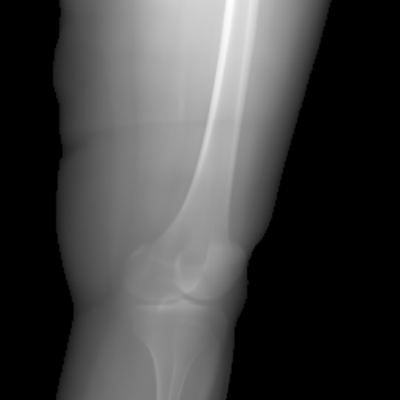}} 
    \centerline{(c2)}\medskip
\end{minipage}
\hfill
    \begin{minipage}[b]{0.18\linewidth}
    \centering
    \centerline{\includegraphics[width=1.6cm]{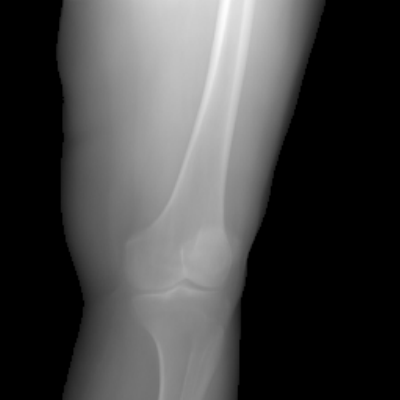}}
    \centerline{(c3)}\medskip
\end{minipage}
\hfill
    \begin{minipage}[b]{0.18\linewidth}
    \centering
    \centerline{\includegraphics[width=1.6cm]{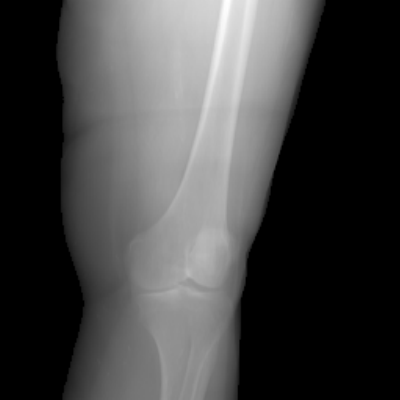}}
    \centerline{(c4)}\medskip
\end{minipage}
    \begin{minipage}[b]{0.18\linewidth}
    \centering
    \centerline{\includegraphics[width=1.6cm]{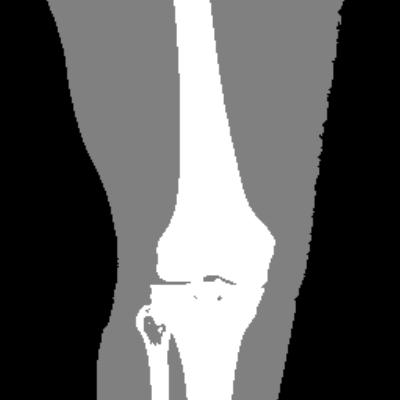}}
    \centerline{(d0)}\medskip
\end{minipage}
\hfill
    \begin{minipage}[b]{0.18\linewidth}
    \centering
    \centerline{\includegraphics[width=1.6cm]{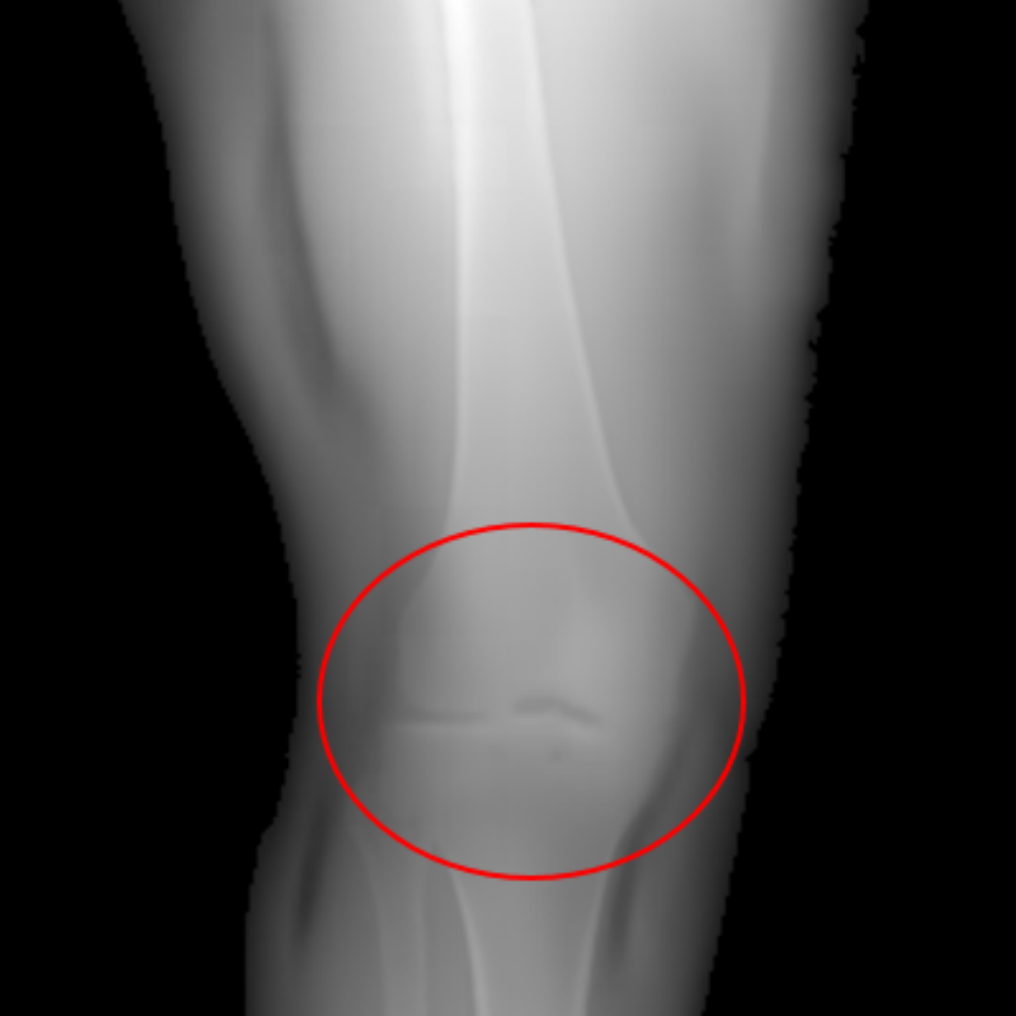}} 
    \centerline{(d1)}\medskip
\end{minipage}
\hfill
    \begin{minipage}[b]{0.18\linewidth}
    \centering
    \centerline{\includegraphics[width=1.6cm]{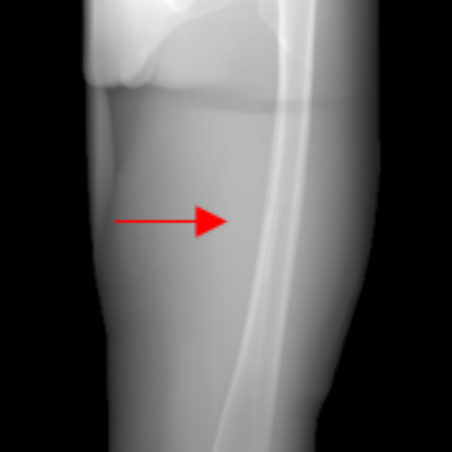}} 
    \centerline{(d2)}\medskip
\end{minipage}
\hfill
    \begin{minipage}[b]{0.18\linewidth}
    \centering
    \centerline{\includegraphics[width=1.6cm]{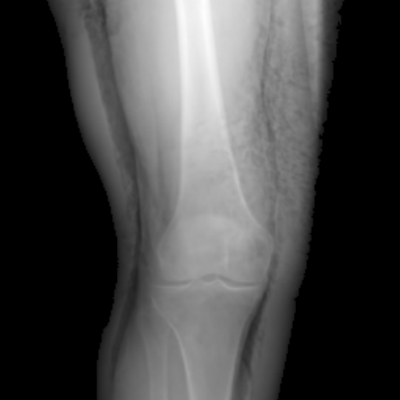}}
    \centerline{(d3)}\medskip
\end{minipage}
\hfill
    \begin{minipage}[b]{0.18\linewidth}
    \centering
    \centerline{\includegraphics[width=1.6cm]{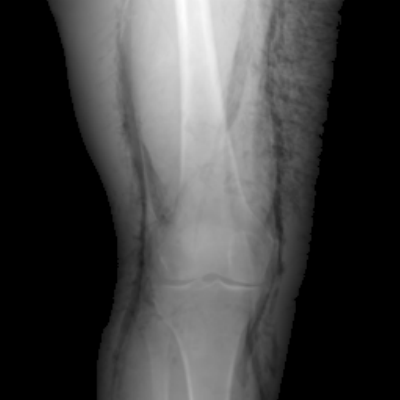}}
    \centerline{(d4)}\medskip
\end{minipage}

    \begin{minipage}[b]{0.18\linewidth}
    \centering
    \centerline{\includegraphics[width=1.6cm]{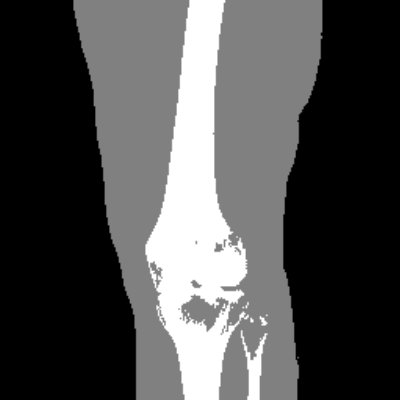}}
    \centerline{(e0)}\medskip
\end{minipage}
\hfill
    \begin{minipage}[b]{0.18\linewidth}
    \centering
    \centerline{\includegraphics[width=1.6cm]{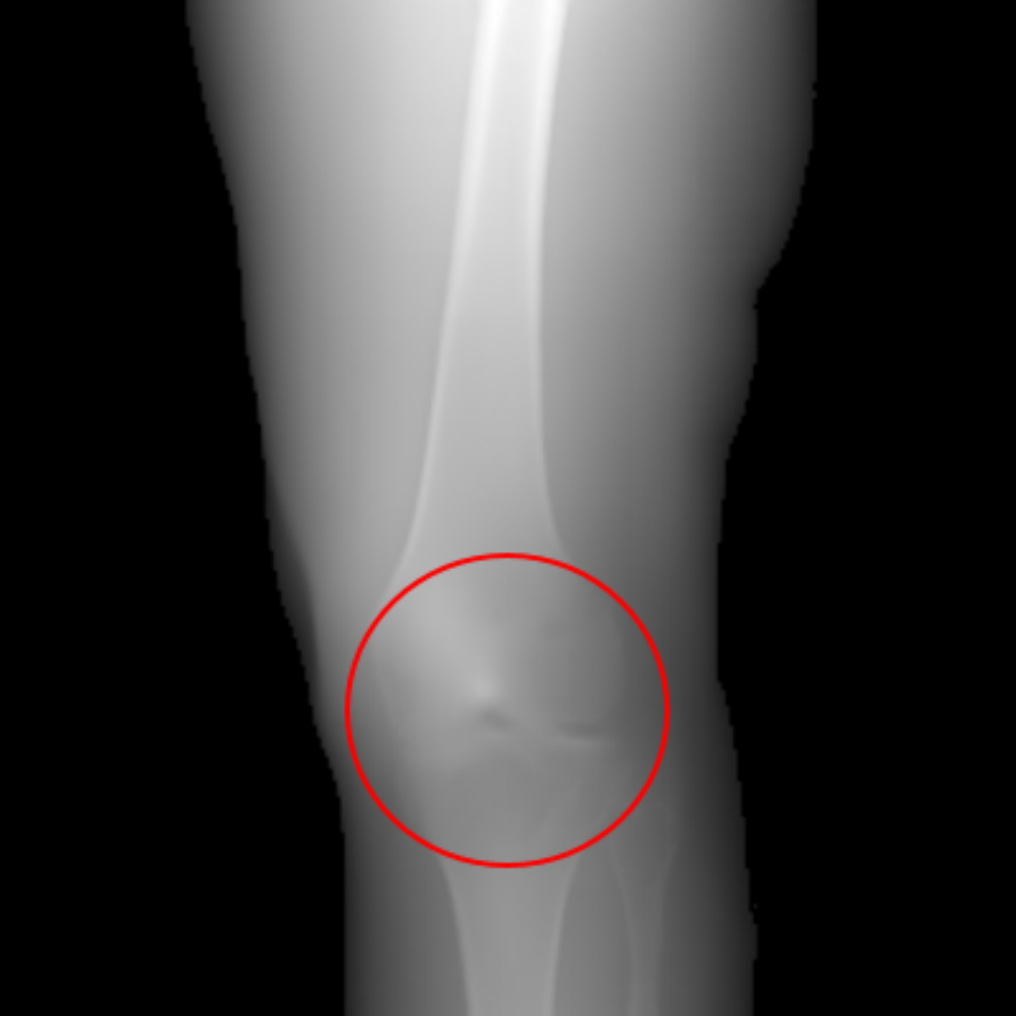}} 
    \centerline{(e1)}\medskip
\end{minipage}
\hfill
    \begin{minipage}[b]{0.18\linewidth}
    \centering
    \centerline{\includegraphics[width=1.6cm]{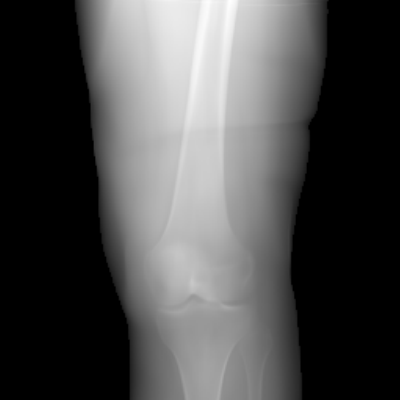}} 
    \centerline{(e2)}\medskip
\end{minipage}
\hfill
    \begin{minipage}[b]{0.18\linewidth}
    \centering
    \centerline{\includegraphics[width=1.6cm]{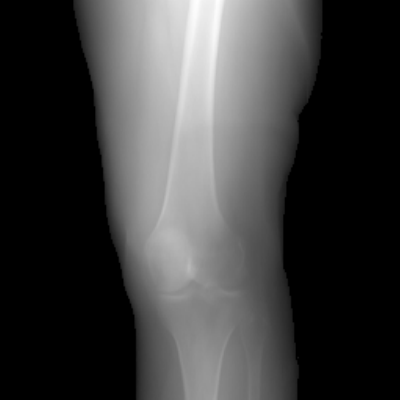}}
    \centerline{(e3)}\medskip
\end{minipage}
\hfill
    \begin{minipage}[b]{0.18\linewidth}
    \centering
    \centerline{\includegraphics[width=1.6cm]{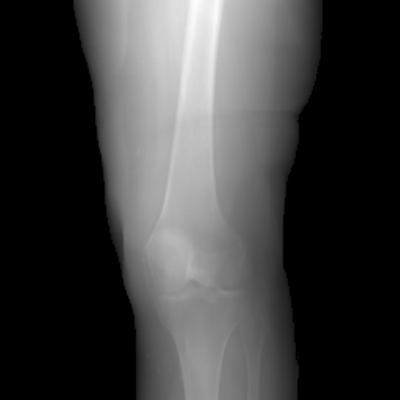}}
    \centerline{(e4)}\medskip
\end{minipage}

    \begin{minipage}[b]{0.18\linewidth}
    \centering
    \centerline{\includegraphics[width=1.6cm]{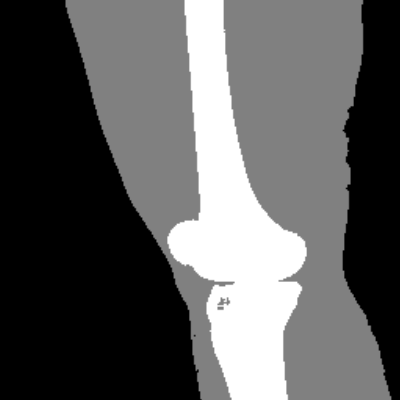}}
    \centerline{(f0)}\medskip
\end{minipage}
\hfill
    \begin{minipage}[b]{0.18\linewidth}
    \centering
    \centerline{\includegraphics[width=1.6cm]{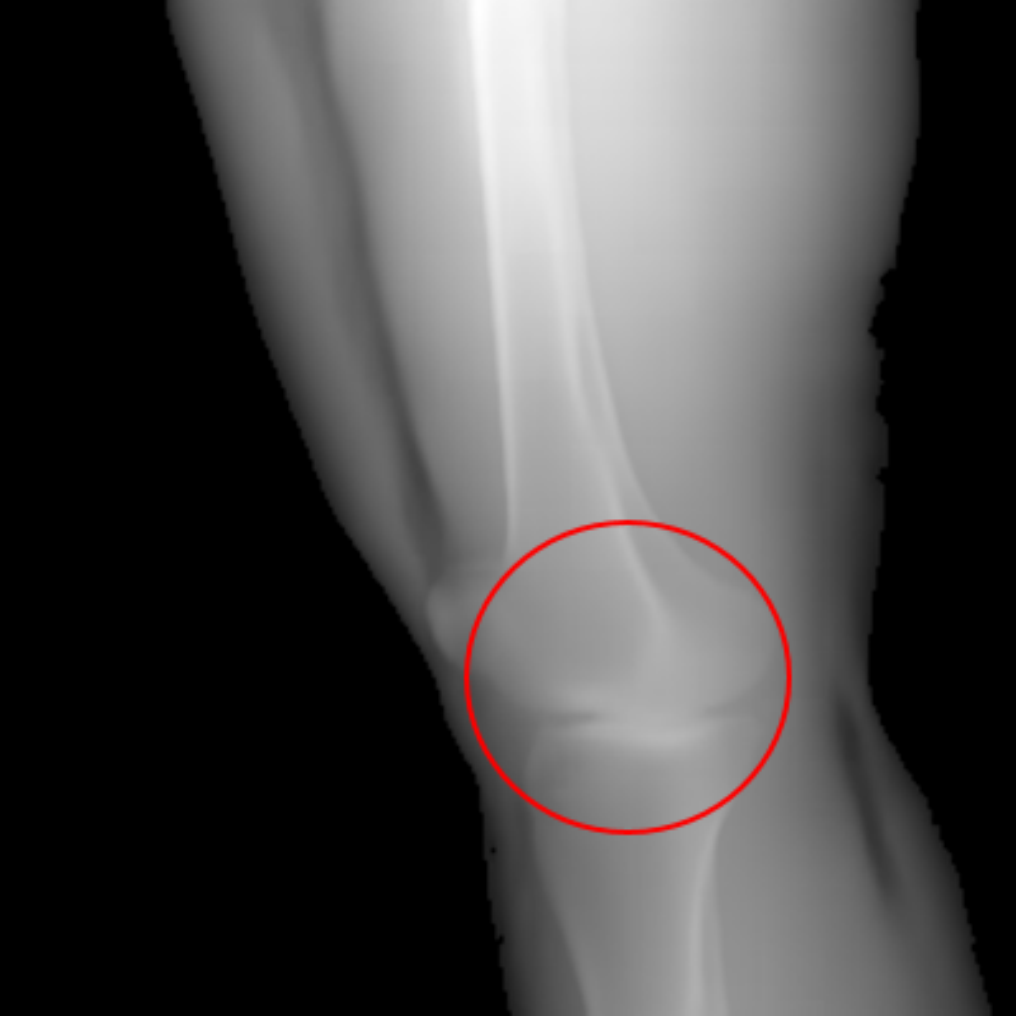}} 
    \centerline{(f1)}\medskip
\end{minipage}
\hfill
    \begin{minipage}[b]{0.18\linewidth}
    \centering
    \centerline{\includegraphics[width=1.6cm]{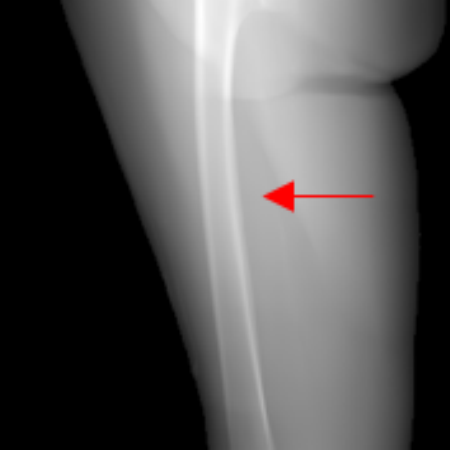}} 
    \centerline{(f2)}\medskip
\end{minipage}
\hfill
    \begin{minipage}[b]{0.18\linewidth}
    \centering
    \centerline{\includegraphics[width=1.6cm]{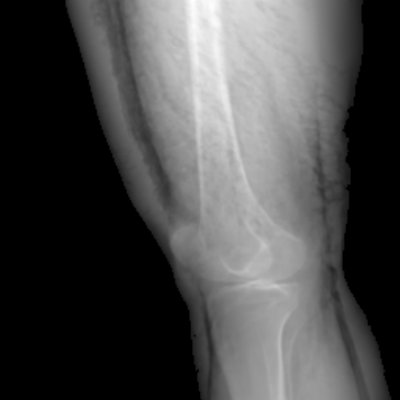}}
    \centerline{(f3)}\medskip
\end{minipage}
\hfill
    \begin{minipage}[b]{0.18\linewidth}
    \centering
    \centerline{\includegraphics[width=1.6cm]{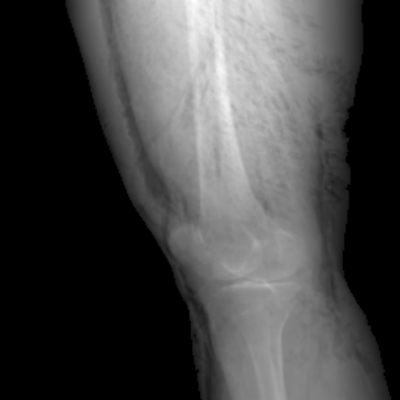}}
    \centerline{(f4)}\medskip
\end{minipage}
\caption[figure3]{\label{fig:figure3} The generated samples under different conditions. Rows (a)-(c): contour segmentation. Rows (d)-(f): contour and bone segmentation.}
\end{figure}

In addition, two different segmentations for each DRR were automatically generated as follows. The first segmentation extracted the leg contour using a threshold of 0.1, having a value of 0.5 for the contour and 0 for the background (refer to a(0)-c(0) in Fig.~\ref{fig:figure3}). The second segmentation extracted bones by thresholding the original CT volume, followed by forward projecting the bone. The bone projection was set to 0.5, and adding it to the contour segmentation resulted in the second segmentation, with bones having a value of 1 and contour a value of 0.5 (refer to d(0)-f(0) in Fig.~\ref{fig:figure3}). In total, 3300 radiographs are generated for each type of segmentation. They were randomly split into a 9:1:1 ratio for training, validation, and testing.

\subsection{Network Structure and Hyperparameters}
\label{sec: network}

The backbone of the neural network employs the noise-conditional score network++ (NCSNpp)~\cite{song2020score}. As illustrated in Fig.~\ref{fig:figure2}, we configure six resolution levels of $(256,128,64,32,16,8)$ with a corresponding number of channels of $(64, 128, 128, 128, 128, 256)$. Furthermore, the time-conditional noise scale $\sigma_t$ is encoded to random Gaussian feature~\cite{tancik2020fourier} and embedded into all residual blocks. Importantly, for CSM only the perturbed X-ray images are input to the network. When conducting the CTM, additional segmentations are concatenated in the input. To compare the diffusion models with naive image-to-image models, the noise-encoding module is removed to form an improved U-Net model and then trained using the $L1$ loss. 






In our experiments, the parameters $\sigma_{min}$ and $\sigma_{max}$ are set to 0.01 and 128. We use a batch size of 16 and the Adam optimizer with learning rate $2\times 10^{-4}$ for training. During the sampling process, a Langevin dynamic corrector with a signal-to-noise ratio of 0.4 is supplied after reverse SDE at each sampling step, and the number of sampling steps $N$ is chosen as 500 to improve sampling speed. Moreover, the hyperparameter $t_0$ for CSM is set to 0.4. All models are trained on four Nvidia A100 GPUs on a single cluster node with a cap of 300 epochs.

\section{Results and Discussion}

In this section, we provide qualitative and quantitative results of the two proposed diffusion-based methods and compare them with the baseline U-Net model. The first column of Fig.~\ref{fig:figure3} showcases six randomly selected conditions: (a0)-(c0) show contour segmentations, and (d0)-(f0) denote segmentations containing contour and bones. In Fig.~\ref{fig:figure3} (a1)-(f1), the images generated by U-Net contain blurred fine details in locations where bones overlap, despite maintaining the given shape, as highlighted by the red circle. In contrast, the results from CSM appear more realistic than the U-Net. However, their quality decreases with introduced constraints, as indicated by the red arrow in Fig.~\ref{fig:figure3} (d2) and (f2). The results from CTM not only achieve nearly the same level of fineness as the labels but also provide reasonable results with respect to the given conditions as illustrated in the fourth column.

\begin{table}[htb]
    \centering
    \begin{tabular}{c|c|c|c}
        \hline
         \multicolumn{2}{c|}{Condition}&  Contour & Contour+bone\\
         \hline
         \multirow{2}{*}{U-Net} & MAE & $0.0209{\pm 0.007}$ & $0.0188{\pm 0.006}$ \\
         \cline{2-4}
         & PSNR (dB) & $29.188{\pm 2.22}$ & $30.304{\pm 2.45}$ \\
         \cline{2-4}
         \hline
         \multirow{2}{*}{CSM} & MAE & $0.0395{\pm 0.010}$ & $0.0507{\pm 0.010}$ \\
         \cline{2-4}
         & PSNR (dB) & $22.911{\pm 1.89}$ & $21.350{\pm 1.49}$ \\
         \cline{2-4}
         \hline
         \multirow{2}{*}{CTM} & MAE & $\mathbf{0.0193{\pm 0.005}}$ & $\mathbf{0.0152{\pm 0.007}}$ \\
         \cline{2-4}
         & PSNR (dB) & $\mathbf{29.498{\pm 1.91}}$ & $\mathbf{31.680{\pm 1.76}}$ \\
         \cline{2-4}
         \hline
    \end{tabular}
    \caption{Quantitative model comparison.}
    \label{tab:my_label}
\end{table}

Table~\ref{tab:my_label} summarizes the quantitative results averaged across all testing data. The evaluation metrics include mean absolute error (MAE) and peak signal-to-noise ratio (PSNR). We observed that CTM performs substantially better than U-Net and CSM under both segmentation-based conditions, and CSM performs worse than the U-Net. 

Unlike the U-Net which learns a mapping function between input and output, the diffusion models can implicitly capture the underlying data distribution from the training data and then sample it, preventing the loss of fine details on the pixel level. However, in CSM, conditions are incorporated only at the first sampling step while being perturbed, which results in imprecise conditional information. Instead, CTM provides an estimated score function of the conditional distribution for each sampling step, accommodating both reliability and realism. Nonetheless, presently generated X-ray images only encompass independent 2D conditional information, which may introduce geometric inconsistencies between a set of projections. Future research will focus on modeling 3D probabilistic distributions with the provided 2D conditions to enable CT reconstruction from the generated projections. In addition, clinical datasets will also be incorporated.

\section{Conclusion}

In this work, we explored two different pipelines of diffusion models to generate segmentation-conditioned knee X-ray data. The results demonstrate that both methods can generate realistic radiographs under the given conditions, with the method of conditional training achieving more stable performance. Ultimately, these high-quality synthetic medical images have the potential to benefit the development of data-driven research and educational applications in the medical field.


\section*{Acknowledgment}

This work was supported by the European Research Council (ERC Grant No. 810316). The authors gratefully acknowledge the HPC resources provided by NHR@FAU using hardware funded by the German Research Foundation (DFG).

\ifCLASSOPTIONcaptionsoff
  \newpage
\fi

\bibliographystyle{IEEEtran}
\bibliography{ref}

\end{document}